\documentclass[12pt,preprint]{aastex}

\shorttitle{Class Discovery}

\begin{document}
\title{Class Discovery in Galaxy Classification}
\author{David Bazell}
\affil{Eureka Scientific, Inc., 6509 Evensong Mews, Columbia, MD
  21044}
\email{bazell@comcast.net}
\and
\author{David J. Miller}
\affil{Department of Electrical Engineering, Pennsylvania State
  University}
\email{millerdj@ee.psu.edu}

\begin{abstract}
In recent years, automated, supervised classification techniques have
been fruitfully applied to labeling and organizing large astronomical
databases.  These methods require off-line classifier training, based
on labeled examples from each of the (known) object classes.  In
practice, only a small batch of labeled examples, hand-labeled by a
human expert, may be available for training.  Moreover, there may be
{\it no} labeled examples for some classes present in the data, i.e.
the database may contain several {\it unknown classes}.  Unknown
classes may be present due to 1) uncertainty in or lack of knowledge
of the measurement process, 2) an inability to adequately ``survey'' a
massive database to assess its content (classes), and/or 3) an
incomplete scientific hypothesis.  In recent work, new class discovery
in mixed labeled/unlabeled data was formally posed, with a proposed
solution based on mixture models.  In this work we investigate this
approach, propose a competing technique suitable for class discovery
in neural networks, and evaluate both methods for classification and
class discovery on several astronomical data sets.  Our results
demonstrate up to a $57\%$ reduction in classification error
compared to a standard neural network classifier that uses only
labeled data.
\end{abstract}


\section{Introduction}
Finding patterns in large astronomical databases and grouping the data
into different classes has become an important task in recent years,
one that must be done in an automated fashion given the massive
amounts of sky survey data currently being collected and stored.
Traditional methods such as looking at large sky plates and
identifying galaxies and clusters by eye are no longer feasible.
Within statistical pattern recognition there are two traditional
approaches to data classification: supervised statistical
classification and unsupervised learning (clustering).  In the
supervised approach one is given a batch of training data containing
labeled examples from each of the known classes of interest.  These
examples are used to learn a decision function which partitions the
feature space into disjoint regions, each associated with one of the
classes.  Typical decision function structures used in practice are
neural networks, decision trees, and prototype-based classifiers.
Once the decision function is learned it can be used to automatically
classify new examples.  The supervised learning of the decision
function can be slow and generally requires off-line training.
Moreover, enough labeled examples from each of the
classes are required to learn an accurate decision function that
adequately separates data into the different classes.  However,
extracting labeled examples from a large database is a time consuming
and expensive process, generally requiring hand labeling by human
experts.

Alternatively, unsupervised learning, or clustering, techniques assign
data to groups without any need for supervising examples.  In these
approaches, the grouping is chosen so that the data examples belonging
to each cluster are ``as similar as possible'' and so that examples
from different clusters are ``as dissimilar as possible''.  This
notion of similarity is quantified through a mathematical clustering
objective function, one which relies on the choice of a distance
measure defined on the feature space, e.g. the sum-of-squared-errors
criterion.  Mixture models \citep{Duda, Mclachlan, Raftery} are one
form of model-based clustering.  They produce probabilistic, or soft,
assignments of data points to each of the mixture components, or
clusters.  The nature and quality of the learned groupings obtained
via unsupervised clustering critically depends upon the choice of
clustering distance measure and also on the number of clusters to be
learned, which must be specified as part of the algorithm.  There are
currently no generally agreed upon approaches for choosing these
parameters in unsupervised learning.  Furthermore, without supervising
examples, there are no guarantees that the chosen parameters are
consistent with the learning of clusters that correspond to the ground
truth classes in the data.

In recent years, seeking to overcome the disadvantages of both
supervised and unsupervised learning, {\it semisupervised learning}
techniques have been proposed, e.g. \citet{Shashahani},
\citet{Miller1}, \citet{Nigam}.  These methods learn based on a batch
of data that consists of both labeled {\it and} unlabeled examples.
On the one hand, appropriate use of unlabeled examples, in addition to
labeled ones, can help to better learn the ``shapes'' of each of the
classes (i.e., the class-conditional density functions)
\citet{Shashahani}, \citet{Miller1}.  On the other hand, use of some
labeled examples can potentially help to guide unsupervised clustering
methods toward solutions that capture the ground truth classes in the
data \citet{Miller1}, \citet{Basu1}.

In nearly all prior semisupervised work it has been assumed that the
number of classes present in the database is known and that there are
labeled examples from each of these classes.  However, for scientific
domains, especially those with massive data collections, this
assumption may not be very reasonable.  There are several reasons why
the presence of some classes within a given data set may be unknown.
First, there may be uncertainty associated with the measurement
process.  As one example, suppose the data was measured by a new
device or one whose operation (e.g., measurement sensitivity) is
imprecisely known.  If the device's sensitivity or dynamic range is
greater than was supposed, it may record measurements corresponding to
unanticipated events or objects.  Second, in some cases, the set of
known classes is inferred by surveying or sampling a subset of the
collected database.  However, if there are millions of data samples,
it is only practical to sample a very small data subset -- if 99 \% of
a database remains unsurveyed, it is quite possible that important
content (such as some classes) may be missed.  Finally, the set
of known classes reflects the currently accepted scientific hypotheses
for a given domain.  Unknown classes may be present in the data
because the current theory is wrong or incomplete.  In fact, we may go
so far as to say that the assumption that a collected database is
composed of a fixed (known) set of classes is in some way inconsistent
with the scientific method -- one is guaranteed to find what one is
looking for, i.e., known classes, rather than what may actually be
present in the data.

In recent work \citet{pami}, \citet{Browning} the problem of new class
discovery in mixed labeled/unlabeled data sets was formally proposed.
The authors recognized that, within a mixed labeled/unlabeled data
set, unknown classes will consist of clusters or groups that are {\it
purely unlabeled}.  The authors proposed a special mixture modeling
technique tailored for discovering the cluster/group structure in the
data and, in particular, the unknown classes.  In their approach,
individual mixture components either represent data from known
classes, in which case they own some labeled samples, or they
represent unknown classes, in which case they own purely unlabeled
data subsets.  Their learning approaches were demonstrated to be very
effective at identifying purely unlabeled clusters \citet{pami} or
{\it nearly} purely unlabeled ones \citet{Browning} in partially
labeled data sets.  Such clusters represent {\it putative} unknown
classes.  Their approach was further demonstrated to improve the
overall accuracy of the mixture modeling solution.

In this work, we consider the problem of galaxy classification, based
on sky survey data, with several unknown classes present in the data.
For this domain, we evaluate both \citet{Browning} and a new approach
which we propose here, one that is applicable to class discovery for
neural network-based (NN) classifiers.  In section 2, we describe the
data sets and data preparation.  In section 3, we review
\citet{pami}, \citet{Browning} and also introduce a class discovery
approach for NN classifiers.  In section 3, we also describe several
performance criteria, each capturing different aspects of the class
discovery problem.  In section 4, we present our experimental results.
Finally, the paper concludes with a summary and some discussion.

\section{Data Preparation}
In our experiments we used two data sets, each with over 5000 data
points. The first was data from \citet{storrie92} (henceforth denoted
as ESOLV after the ESO-LV catalog of \citet{lauberts89}) which has
been used previously in several studies of automated classification
methods \citep{storrie92, owens96, bazell01}.  The second data set
consisted of SDSS early release data \citet{stoughton}, composed of
over 50000 objects of various types.

\citet{storrie92} performed one of the earliest attempts at
morphological classification of galaxies using neural networks.  Their
data set consisted of 13 input features derived from images of
galaxies which were then used to classify the galaxies into five
classes: E, S0, Sa + Sb, Sc + Sd, and Irr.  We used their input data
set of 5217 galaxies.  The features in this data set are described in
\citet{storrie92}.  \citet{bazell01} describes the use of this data
set for galaxy classification using ensembles of neural networks.  For
our studies we eliminated one of the features, $E^{Fit}_{Err}$, which
is the error in an ellipse fit to B isophotes.  This feature had very
small variance and equaled zero for approximately 80\% of the objects.
Thus, we used 12 of the 13 features in the original data set.

We also used a data set with an order of magnitude more objects than
the ESOLV data.  The SDSS data consists of 54007 objects drawn from
seven different classes.  Each object is described by a total of six
features: photometric values in u, g, r, i, and z, and the redshift of
each object.

Tables \ref{tab:esolv} and \ref{tab:sdss} summarize the properties of
the data sets we used.  For each class the tables show the number of
objects in the class, the percentage of total objects that class
represents, and the type of object in the class.

\begin{deluxetable}{crrl}
\tabletypesize{\scriptsize}
\tablecaption{ESOLV dataset summary \label{tab:esolv}}
\tablewidth{0pt}
\tablehead{
\colhead{Class} & \colhead{Number} & \colhead{\%} & \colhead{Object}}
\startdata
0 & 466    &  8.93 & E \\
1 & 851    & 16.3  & S0 \\
2 & 2403   & 46.1  & Sa + Sb \\
3 & 1132   & 21.7  & Sc + Sd \\
4 & 365    &  7.00 & Irregular \\
\enddata
\end{deluxetable}

\begin{deluxetable}{crrl}
\tabletypesize{\scriptsize}
\tablecaption{SDSS dataset summary \label{tab:sdss}}
\tablewidth{0pt}
\tablehead{
\colhead{Class} & \colhead{Number} & \colhead{\%} & \colhead{Object}}
\startdata
0 & 229    &  0.42 & Unknown Spectrum \\
1 & 6049   & 11.2  & Stellar Spectrum \\
2 & 41930  & 77.6  & Galaxy Spectrum \\
3 & 4409   &  8.16 & Quasar Spectrum \\
4 & 237    &  0.44 & High z Quasar Spectrum \\
5 & 130    &  0.24 & Sky Spectrum \\
6 & 1023   &  1.89 & Late Type Star \\
\enddata
\end{deluxetable}

For our experiments, we treated one or two of the classes as being
unknown, withholding from use during model learning the label
for every data example from each of the unknown classes.  For data
from all other classes, we retained the labels for a randomly selected
subset (roughly 10\% of the points from these classes--521 examples
for the ESOLV data and 5400 examples for the SDSS data).  The random
selection was performed in a ``stratified'' fashion, ensuring that the
number of labeled examples from each known class is in proportion to
the mass, or frequency of occurrence, of the class.  In this way, we
obtained a data set containing both labeled and unlabeled examples,
and with all labels missing from one or two classes.  This is
precisely the data scenario proposed and addressed in \citet{pami} and
\citet{Browning}.

\section{Description of Algorithms}
We used two algorithms to classify the data and perform class
discovery, a mixture model and a backpropagation algorithm.  These
approaches are described in detail below.

\subsection{Mixture Modeling Approach}
This subsection reviews the work in \citet{pami}, \citet{Browning}.
There are three main contributions in these works: 1) the problem of
new class discovery in mixed labeled/unlabeled data was proposed; 2) a
mixture model was proposed for this scenario, one that incorporates a
realistic labeling mechanism.  This model has built into it the
competing hypotheses that a data sample may come from known or
unknown/outlier groups.  Thus, this model naturally yields {\it a
posteriori} probabilities for these hypotheses, as well as the
standard {\it a posteriori} probabilities on the known classes (now
conditioned on a known class hypothesis);  3) methods for learning the
mixture model from given data were proposed in
\citet{pami}, \citet{Browning}.  In these approaches, individual mixture
components learn to represent either known or unknown class
data.  We next review \citet{pami}, \citet{Browning} in more detail.

Consider a data set ${\cal X}_m = \{{ \cal X}_l,{\cal X}_u\}$, where
${\cal X}_l = \{(\underline{x_1},c_1), (\underline{x_2},c_2), \ldots,$
$(\underline{x_{N_l}},c_{N_l})\}$ is the labeled subset and ${\cal
X}_u = \{\underline{x_{N_l+1}}, \ldots, \underline{x_{N}}\}$ is the
unlabeled subset.  Here, $\underline{x}_i \equiv
(x_{i1},x_{i2},\ldots,x_{id})$ is a feature vector and $c_i$ is a
class label from the set of known classes ${\cal P}_c$.  This mixed
data scenario was considered previously in e.g.  \citet{Shashahani},
\citet{Miller1}, \citet{Nigam}.  Unlike these works, a key element in
\citet{pami}, \citet{Browning} summarized here, is that the fact that
a sample is unlabeled is treated {\it as observed data}.  Accordingly,
we redefine ${\cal X}_m = \{{\cal X}_l,{\cal X}_u\}$, where now ${\cal
X}_l = \{(\underline{x_1},{\rm `l'},c_1), (\underline{x_2},{\rm `l'
},c_2), \ldots, (\underline{x_{N_l}},{\rm `l'},c_{N_l})\}$ and ${\cal
X}_u = \{(\underline{x_{N_l+1}},{\rm `m'}), \ldots, (\underline{x_{
N}},{\rm `m'})\}$.  Here the new random observation ${\cal L } \in
\{{\rm `l'},{\rm `m'}\}$ is introduced, taking on values indicating a
sample is either labeled or missing the label.  If a sample is
labeled, then it is known the sample originates from one of the known
classes.  On the other hand, if the sample is unlabeled, then there
are two sources of uncertainty.  First, under the assumption that
there may be unknown classes present, it is unknown whether or not the
given sample originates from a known class.  Second, conditioned on
its belonging to one of the known classes, the class of origin for the
sample is unknown.  In \citet{pami}, a mixture model was proposed that
explains all the observed data, including the presence or absence of a
label.  Two types of mixture components were posited, differing in the
mechanism they use for generating label presence/absence.
``Predefined'' components generate both labeled and unlabeled data and
assume labels are missing at random.  These components represent the
known classes.  ``Non-predefined'' components only generate unlabeled
data -- thus, in localized regions, they capture data subsets that are
purely unlabeled.  {\it Such subsets may represent an outlier
distribution or new classes}.  An example is shown in Figure 1, with
labeled data denoted by a single number, the class, and with unlabeled
data denoted by `U' followed by the ground truth class of origin.

\begin{figure}
\plotone{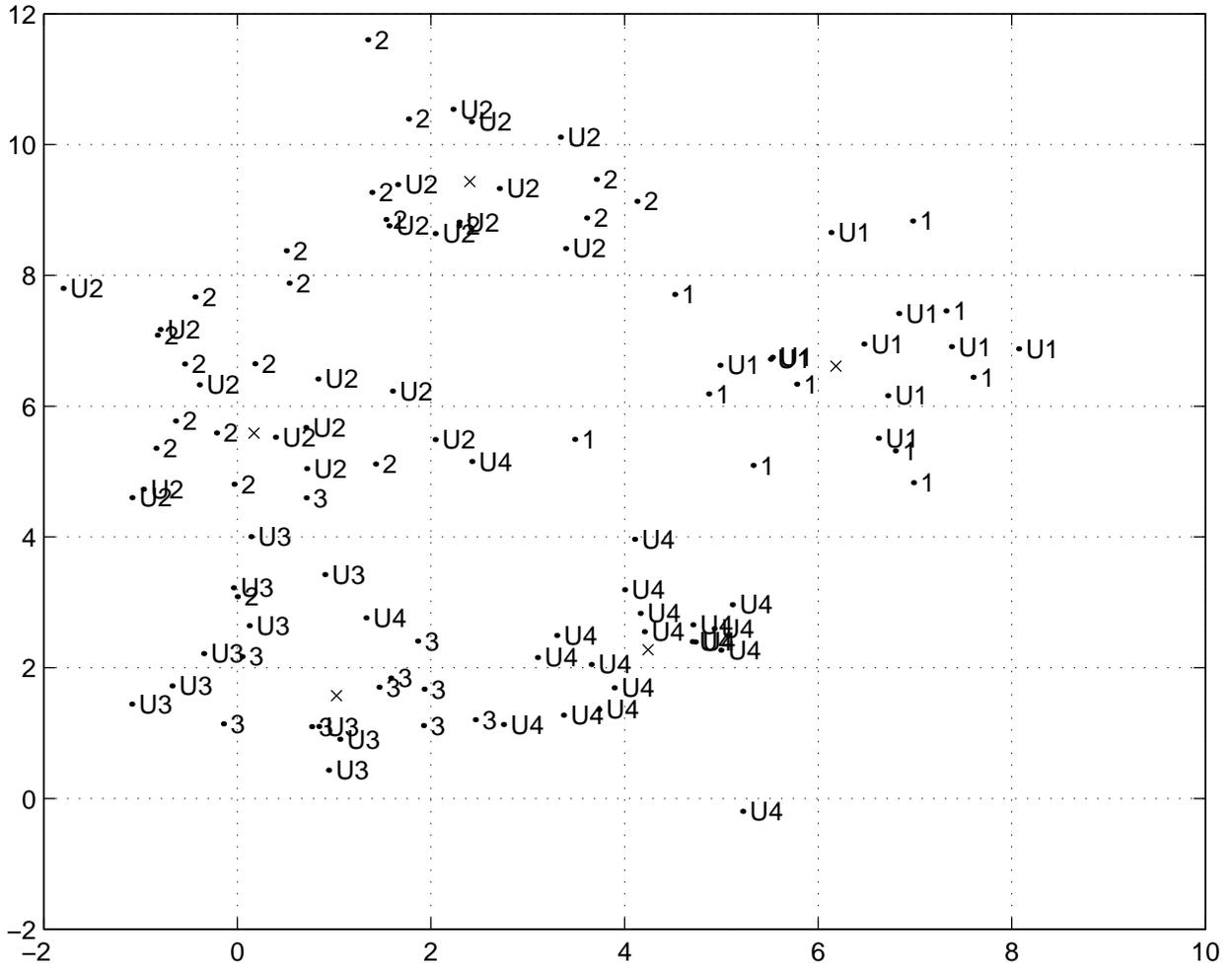}
\label{fig:fig1}
\caption{\small Example involving partially labeled data and an
unknown class.  Labeled are denonted by a single class number while
unlabeled data are denoted by a `U' followed by the class.  The $x$s
denote mixture component centers.}
\end{figure}
The 2-D data points were generated according to a Gaussian mixture
with 5 components.  For this example, all points originating from the
same component come from the same class, i.e. classes ``own'' either
one or multiple mixture components.  In this example, class two
consists of two components, with the other classes consisting of
single components.  Note that for the known classes, it appears in
Fig. 1 (and is the case) that labels are missing at random, while for
the unknown classes (class 4 in this case), labels are
deterministically (always) missing.  This is consistent with an expert
labeler who randomly selects a subset of data from the known classes
and is unable to label any data from unknown classes.  In
\citet{pami} and \citet{Browning}, mixture models were proposed that were
tailored to this data scenario.  In this work, we have applied the
approach in \citet{Browning}, summarized next.

\vspace{0.2in}
\noindent
{\it Notation}

\vspace{0.2in} Let ${\cal M}_k$, $k=1,\ldots,M$ denote the $k$th
mixture component.  Let ${\cal C}_{\rm pre}$ denote the subset of
``predefined'' components, with the remaining subset denoted
${\bar{\cal C}}_{\rm pre}$.  Let $C \in {\cal P}_c \equiv
\{1,2,\ldots,N_c\}$ be a random variable defined over the $N_c$ known
classes, with $c(\underline{x}) \in {\cal P}_c$ the class label paired
with $\underline{x}$.  Let $\alpha_k$ denote the prior probability for
component $k$, $\theta_k$ the parameter set specifying component $k$'s
(component-conditional) joint feature density, and let
$P[\underline{x} | \theta_k ]$ denote this density.  We also introduce
a new class set $\tilde{\cal P}_c = \{1,2,\ldots,N_c,u\}$, consisting
of the set ${\cal P}_c$ plus a value $u$, used to indicate that a
sample is unlabeled.  With respect to the class set $\tilde{\cal
P}_c$, every sample is now labeled, with the unlabeled samples taking
on the label values `u'.  We suppose a different random label
generator, conditioned on each mixture component, i.e. ${\rm Prob}[c
|{\cal M}_k] \equiv \beta_{c|k}, c \in \tilde{\cal P}_c$, where
$\sum\limits_{c \in \tilde{\cal P}_c} \beta_{c|k} = 1$.  In summary,
the model is based on the parameter set $\Lambda =
\{\{\alpha_k\},\{\theta_k\}, \{\beta_{c|k}\}\}$.

\vspace{0.05in}
\noindent
{\bf Hypothesis for Random Generation of the Data}

\noindent
This model \citet{Browning} hypothesizes that each sample from ${\cal
X}_m$ is generated independently, based on $\Lambda$, according to the
following stochastic generation process:

\noindent
i) Randomly select a component ${\cal M}_j$ according to $\{\alpha_k\}$.

\noindent
ii) Randomly select a vector $\underline{x}$ according to
$P(\underline{x} | \theta_j)$ and a label $c$ according to
$\{\beta_{k|j}\}$.

\vspace{0.05in}
\noindent
{\bf Joint Data Likelihood}

\noindent
The log of the joint data likelihood associated with this model is
\begin{eqnarray}
\label{newlik}
{\cal L} = \sum\limits_{\underline{x} \in {\cal X}_m}
\log\sum\limits_{k =1}^M \alpha_k f(\underline{x} | \theta_k)
\beta_{c(\underline{x}) | k},
\end{eqnarray}
where $c(\underline{x}) \in \tilde{\cal P}_c$.  The model parameters
$\Lambda$ can be chosen to maximize the log-likelihood (\ref{newlik})
via the Expectation-Maximization (EM) algorithm (e.g. \citet{Duda}).
Since the derivation of these EM equations is standard, their
exposition is herein omitted.

This model does not explicitly discover new class components,
i.e. mixture components that are purely unlabeled.  However, suppose
that, for a given component ${\cal M}_j$, we have that $\beta_{u | j}
\simeq 1$ and $\beta_{u|j}$ is also significantly greater than the
{\it average} value $\frac{1}{M} \sum\limits_{j'=1}^M \beta_{u|j'}$.
In this case, the fraction of unlabeled data owned by the component is
unusually high.  We categorize these components as ``nonpredefined'',
i.e. ${\cal M}_j \in {\bar{\cal C}}_{\rm pre}$.  Such components are
putative unknown class components.  All other components are
categorized as ``predefined'', representing known class data.  To
summarize, we have the following strategy for class discovery/outlier
detection in mixed data: 1) learn a mixture model to maximize the log
likelihood (\ref{newlik}); 2) for each component, declare it
``nonpredefined'' if $\beta_{u | j} - (\frac{1}{M}
\sum\limits_{j'=1}^M \beta_{u|j'}) > \delta$; otherwise, declare it
``predefined''.  Here, $\delta$ is a suitably chosen threshold.  In
practice, we declare a component ``nonpredefined'' when its value
$\beta_{u|j}$ is closer to $1.0$ than to the average value, i.e., we
choose $\delta = \frac{1}{2}(1 - \frac{1}{M}\sum\limits_{j'=1}^M
\beta_{u|j'})$.  We have found this choice for $\delta$ to give
reasonable results for a variety of experimental conditions (for
different data sets and for different fractions of labeled data).

\vspace{0.2in}
\noindent
{\it Statistical Inferences from the Model}

\vspace{0.2in} After applying this thresholding operation to each
component, the resulting model is naturally applied to address 
several inference tasks: 1) standard classification of a given sample
to one of the known classes; and 2) known vs. unknown
class discrimination.  For classification to known classes, for a
given sample $\underline{x}$, we compute the {\it a posteriori}
probabilities

\begin{eqnarray}
\label{inf1}
P[C=c | \underline{x}; \Lambda] = \frac{\sum\limits_{k \in {\cal
C}_{\rm pre}} \alpha_k f(\underline{x} | \theta_k)
(\frac{\beta_{c|k}}{1 - \beta_{u|k}})}{\sum\limits_{k \in {\cal C
}_{\rm pre}} \alpha_k f(\underline{x} | \theta_k)
}, c \in {\cal P}_c.
\end{eqnarray}
These can be used in a maximum {\it a posteriori} (MAP) class decision rule.

In order to discriminate between the hypotheses that an unlabeled
sample originates from a known versus an unknown class, we need the
{\it a posteriori} probability that the given feature vector is
generated by a nonpredefined component.  This is given by
\begin{eqnarray}
\label{inf2}
P[{\cal M}_{\rm np} | \underline{x} \in {\cal X}_u] =
 \frac{\sum\limits_{k \in {\bar{\cal C}}_{\rm pre}} \alpha_k
 f(\underline {x} | \theta_k) \beta_{u|k}}{\sum\limits_k \alpha_k
 f(\underline{x} | \theta_k) \beta_{u|k}}
\end{eqnarray}

\vspace{0.2in}
\noindent
{\it New Class Discovery}

\vspace{0.2in} While the EM learning assumes that the number of
mixture components $M$ is fixed and known, in practice this size must
be estimated.  Model order selection is a difficult and pervasive
problem, with several criteria proposed \citep{Schwarz, Wallace,
Mclachlan} and no consensus on the right one.  In our class discovery
setting, the importance of accurate model order selection cannot be
overstated -- {\it the nonpredefined components in the validated
solution will be taken as candidates for new classes}, to be forwarded
to a domain expert for further study.  Accurate model order selection
is thus important for successful new class discovery.  Here, as in
\citet{pami}, \citet{Browning} the Bayesian Information Criterion (BIC)
\citep{Schwarz} is applied.  The BIC model selection criterion is
written in the form
\begin{eqnarray}
\label{BIC}
BIC(M) = \frac{N_p(M)}{2} \log N - {\cal L}, 
\end{eqnarray}
with $N_p(M)$ the number of free parameters in the $M$-component model
and $N$ the data length.  The first term is the penalty on model
complexity, with the second term the negative log-likelihood.  We
applied BIC in a ``wrapper-based'' model selection approach; i.e., we
built models for increasing $M$, evaluated each in terms of BIC, and
selected the model with minimum cost.

While any clustering/mixture modeling technique can in principle be
used to discover unknown classes, standard methods do not have any
special impetus for finding label-free (or largely label-free)
clusters.  By contrast, the log likelihood (\ref{newlik}) and the
likelihood function used in \citet{pami} both encourage solutions with
nonpredefined components, when such components are warranted by the
presence of unknown classes in the data.  In (\ref{newlik}), it is the
$\beta_{c|j}$ term which provides the impetus for forming these
unknown classes, since this term approaches its maximum value
($\beta_{u|j}=1$) in the nonpredefined component case.

\subsection{Neural Network Approach}
The mixture modeling approach provides several inference capabilities
when dealing with mixed labeled/unlabeled data sets and possibly
unknown classes: 1) it allows one to infer whether or not a
given sample belongs to one of the known classes; 2) it identifies
purely unlabeled mixture components/clusters, which are reasonably
treated as putative unknown classes or, at any rate, components
of unknown classes; 3) conditioned on a known class hypothesis for a
given sample, the model can infer from which known class the sample
originates (i.e., the usual classification inference capability).

While the mixture modeling approach is naturally suited to new class
discovery given mixed labeled/unlabeled data, neural network (NN)
classifiers do not appear to be predisposed to making these
inferences.  Neural networks are generally trained using a purely
supervised approach, with class labels provided for every example in
the training set.  Thus, in general, unlabeled samples play no role in
the training -- given a mixed labeled/unlabeled data set, the NN
training will discard all the examples from unknown classes, or,
perhaps worse, erroneously impute and use known class labels for this
unknown class data.  Accordingly, the neural network is only
explicitly trained to discriminate between the known classes -- it is
not trained to distinguish known from unknown classes.  While it thus
appears that neural networks do not possess any class discovery
inference capability, we next suggest an approach that give NNs at
least a weak form of this capability.

The neural network algorithm we used is a basic backpropagation
algorithm available with the WEKA machine learning package
\citet{weka}.  We used the default configuration consisting of a three
layer network (input, hidden, and output).  The number of input nodes
was $N_{i}$, one node for each input feature.  There were $N_{c}$
output nodes, one for each known class.  The number of hidden nodes
was calculated according to $N_{h} = (N_{i} + N_{c})/2$.  For the
ESOLV data we used 12 nodes in the input layer corresponding to the 12
input features, eight nodes in the hidden layer, and four in the
output layer.  For the SDSS data we used five input layer nodes, five
hidden layer nodes, and six output layer nodes.

\vspace{0.2in}
\noindent
{\it Decision Confidence}

\vspace{0.2in}
\noindent
Suppose the neural network produces (soft) discriminant function
outputs $g_j(\underline{x})$ for each of the known classes
$j=1,\ldots,N_{c}$.  Let $\tilde{g}_j(\underline{x}) =
\frac{g_j(\underline{x}) - \min_l g_l(\underline{x})}{\sum\limits_k
g_k(\underline{x}) - \min_l g_l(\underline{x})}$.  With this choice,
we have $0 \leq \tilde{g}_j(\underline{x}) \leq 1$ and $\sum\limits_j
\tilde{g}_j(\underline{x}) = 1$, i.e. $\tilde{g}_j(\underline{x})$ is
a probability mass function defined on the known classes.  One
principled measure of uncertainty in these soft decisions is the
Shannon entropy $H = - \sum\limits_j \tilde{g}_j(\underline{x} ) \log
\tilde{g}_j(\underline{x})$.  If $H$ is greater than a preset
threshold, we can declare that the sample $\underline{x}$ does not
convincingly belong to any of the known classes, i.e. it is declared
an unknown class sample.  This approach, based on a measure of the
classifier's degree of indecision, is the one we have taken in
imparting the neural network with some class discovery inference
capability.  Other measures of the classifier's degree of indecision
are also possible.

\subsection{Error Measures}
There are three error measures which we have used to evaluate the
class discovery approaches.  They are defined as follows:

\noindent
{\bf Criterion 1: Misclassification rate in deciding between ``known''
and ``unknown'' class hypotheses}.

\noindent
This criterion was evaluated for both the mixture model and neural
network approaches.  For the mixture model, the classification
decisions were made using a maximum {\it a posteriori} probability
rule, based on (\ref{inf2}).  For the neural network, the decisions
were made based on thresholding of the neural network's entropy
measure, as discussed earlier.  The error rate was measured over the
unlabeled portion of the data set (which consisted of both known and
unknown class data), i.e. it was estimated as the fraction of
unlabeled samples that were misclassified.

\noindent
{\bf Criterion 2: Misclassification rate within ``putative'' unknown classes}

\noindent
The first criterion simply measures how effective an algorithm is at
identifying the subset of (unlabeled) samples that come from new, i.e.
unknown classes.  If there is a single unknown class present in
the data, then this is all that is required.  However, suppose that
there are multiple unknown classes present.  Then, in addition
to identifying the subset of samples from unknown classes, one would
also like to identify the the individual classes which comprise this
unknown class subset.  In other words, one would like to identify the
underlying cluster (group) structure within the unknown class data.
The mixture modeling approach directly models the unknown and,
separately, the known class data by a mixture of components
(clusters).  Each such cluster can be viewed as a putative
unknown class.  A measure of the accuracy of this clustering is the
unknown class label purity of these clusters.  In particular, suppose
one of the learned nonpredefined clusters owns (in a MAP sense) 20
samples that are ground-truth from unknown class A, 30 samples from
unknown class B, and 35 unlabeled samples that in fact belong to known
classes.  The most populous unknown class in the cluster is B.  All
samples in the cluster that do not possess label B are reasonably
counted as errors.  One can sum these errors over all nonpredefined
clusters and divide by the total number of samples owned by all
nonpredefined clusters.  This is the fraction of samples that in
effect have been erroneously assigned to individual nonpredefined
clusters.  Note that this criterion is only well-defined when the
model finds at least one non-predefined component.

\noindent
{\bf Criterion 3: Known class error rate}

\noindent
If a sample belongs to a known class, then we certainly will be
interested in identifying to which known class it belongs.
Accordingly, we can define an error fraction measured over the known
class data.  Several such criteria are possible.  Here, we counted an
error if an unlabeled sample that is from a known class is assigned to
the wrong known class.  For the mixture model, a MAP classification
rule based on the probabilities (\ref{inf1}) was used.

All three of the above criteria require various forms of ground truth
label information for the unlabeled data subset -- Criterion 1
requires a ground truth known/unknown class indication for all the
data, Criterion 2 requires knowledge of unknown class labels for all
the data, and Criterion 3 requires knowledge of all known class
labels.  Since in practice one would not have this information (by
definition, no labels are known for data from unknown classes), these
criteria can only be used for model evaluation/validation.  In
practice, other information sources or expert knowledge would be
needed in order to assess the quality of, or to confirm, the model
inferences.  We also note that the second criterion can only be
evaluated for the mixture model, since the neural network approach
does not attempt to partition the estimated unknown class data into
smaller groups.

\section{Results}
We evaluated the mixture model and neural network on both the ESOLV
and SDSS data sets.  For SDSS, we constrained mixture component
variances to be at least 0.1 in order to avoid an observed tendency
for the learning to find singular solutions (zero variance), as well
as solutions with very small variances along some dimensions.  For
both the mixture model and the neural network there are ``operating
parameters'' whose choices affect the class discovery inference
performance.  For the mixture model, we need to select the model
order, i.e. the number of components.  This choice will clearly affect
the ability to identify unknown class components.  For example, if an
unknown class has a small mass, one or more components will only be
``deployed'' for its representation if the model has many components.
Likewise, if the unknown class has a very large mass (and significant
within-class variation), quite a few components may be needed to
represent it well.  For the neural network, the choice of the entropy
threshold affects performance.  We have performed several experimental
evaluations of the mixture model and neural network, based on
different approaches for choosing these operating parameters.  In
one set of experiments, shown for the ESOLV and SDSS data sets in
Tables \ref{tab:etf} and \ref{tab:stf}, we picked both the mixture
order (over the range 10 to 80) and the NN's entropy threshold (by an
exhaustive search) to maximize Criterion 1 performance.  Note that
these approaches cannot be used in practice since, in performing the
model selection, these methods require evaluating a cost (Criterion 1)
that depends on knowledge of the unknown class labels.  However, this
experiment does allow a comparison of best-case performances achieved
by the mixture and neural network approaches.  For the mixture model,
we have also applied BIC-based selection as described earlier.  This
approach is wholly unsupervised, and thus feasible in practice.

Tables \ref{tab:etf} and \ref{tab:stf} show the results for the ESOLV
data and the SDSS data using the best (lowest) value for Criterion 1.
The first column shows the classes that were treated as unknown for
that series of runs.  The value of ``ncomp'' is the number of
components used by the mixture model corresponding to the best value
of Criterion 1, while ``nonpre'' is the number of nonpredefined
components used by that model.  The remaining columns under ``Mixture
Model'' list error fractions for the three criteria discussed above.
Under ``Neural Network'' we list the value of Criterion 1, the only
error measure evaluated for the neural network.  The last column shows
the percentage change in the Criterion 1 value between the neural
network and the mixture model, with a negative value indicating a
lower Criterion 1 error for the mixture model compared to the neural
network.  The bracketed values at the bottom of each column are the
average values (across all experiments) over that column.  For the
moment we will restrict discussion to the Criterion 1 performance.
Tables \ref{tab:etf} and \ref{tab:stf} show that, with both methods
optimized for Criterion 1 performance, significantly better inference
accuracy is achieved by the mixture-based approach.  For the ESOLV
data we find an average decrease in Criterion 1 error of $20\%$ and a
maximum decrease of $50\%$.  For the SDSS data we find an average
decrease in Criterion 1 error of $57\%$ and a maximum decrease of
$96\%$.  This is not especially surprising, since the mixture is
learned using the unknown class data (but without use of the labels),
while the neural network is only trained on labeled known class data.

In Tables \ref{tab:emdl} and \ref{tab:smdl}, we compare the neural
network, again with the threshold optimized for Criterion 1, against
the mixture model, but with the order now selected based on the BIC
criterion.  Since the neural network decision making threshold is
optimized based on knowledge of the unknown class labels while the
mixture model and its order are chosen without use of this
information, this comparison is not a fair one.  However, in practice
unsupervised order selection will be required.  Thus, this comparison
does give insight into the loss in accuracy attributable to the use
of, generally suboptimal, but practically feasible model order
selection techniques.  As Tables \ref{tab:emdl} and \ref{tab:smdl}
show, the Criterion 1 error for the mixture model is now higher in 5
of 9 cases for the ESOLV data and is the same or higher in 5 of 13
cases for the SDSS data.  For the ESOLV data we find an average
$10\%$ increase in error, while for the SDSS data we find an average
$20\%$ decrease in error, compared to the neural network.

Figure 2 shows a plot of the number of components in the best
performing mixture model as a function of the fraction of objects in
the unknown classes, for the ESOLV data.  There is a clear trend
toward a larger number of components needed to describe data sets
where the unknown classes make up a larger mass fraction of the total
data set.  One point, with classes 0 and 2 as the unknown classes, is
well described by an unexpectedly small number of components (13).
This is discussed below.  A similar trend is not evident in the SDSS
data.  The SDSS data are dominated by two classes (1 and 2) which
together represent over 88\% of the data.

\begin{figure}
\epsscale{0.8}
\plotone{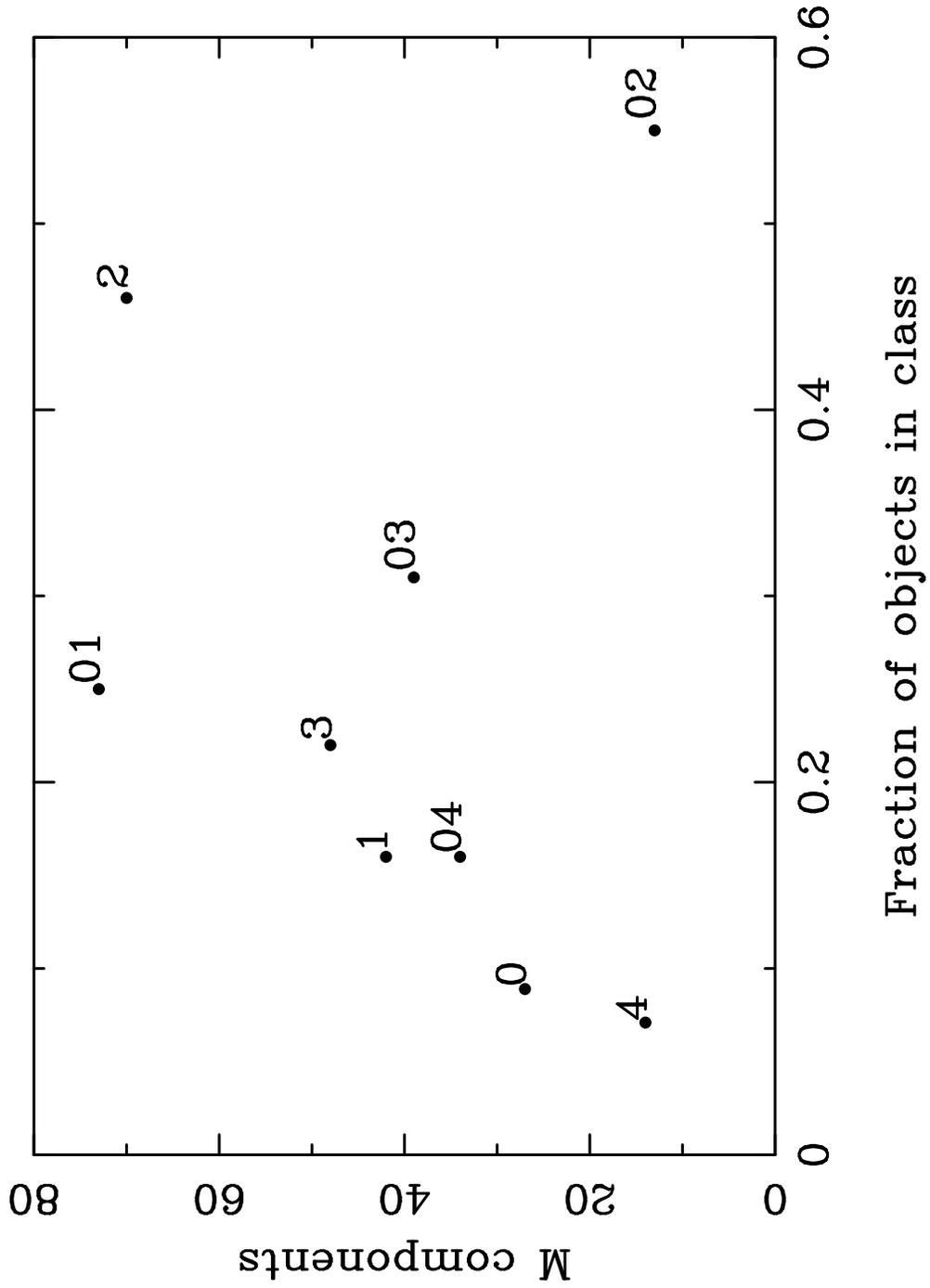}
\label{fig:fig2}
\caption{\small Number of components in the best mixture model, using
  Criterion 1 model selection, as a function of class mass for ESOLV
  data.  Points are labled by unknown class.}
\end{figure}

Figures 3 and 4 show plots of the value of the Criterion 1 error as a
function of the number of components in the mixture model for two
different unknown class combinations.  These plots reflect the
information in Figure 2 in a different way.  When the
unknown classes represent a relatively small mass fraction of the
total number of objects in the data set, the minimum value of
Criterion 1 is found at a relatively moderate number of components.
For example, this is evident in the figure for unknown class 4.
Conversely, as seen in Figure 2, if the unknown classes
comprise a large fraction of the total number of objects in the data
set, then a larger number of components is needed to explain the data
and the minimum value of Criterion 1 is found at a correspondingly
higher number of components.  Note that in Figure 4 Criterion 1
remains static over a range of model orders (for unknown class 0, over
the range 10-20 components).  This is due to the fact that the
performance only changes at discrete points, where additional
nonpredefined components are introduced.  In this case, no
nonpredefined components were introduced until $M$ increased beyond
20.
\begin{figure}
\epsscale{0.8}
\plotone{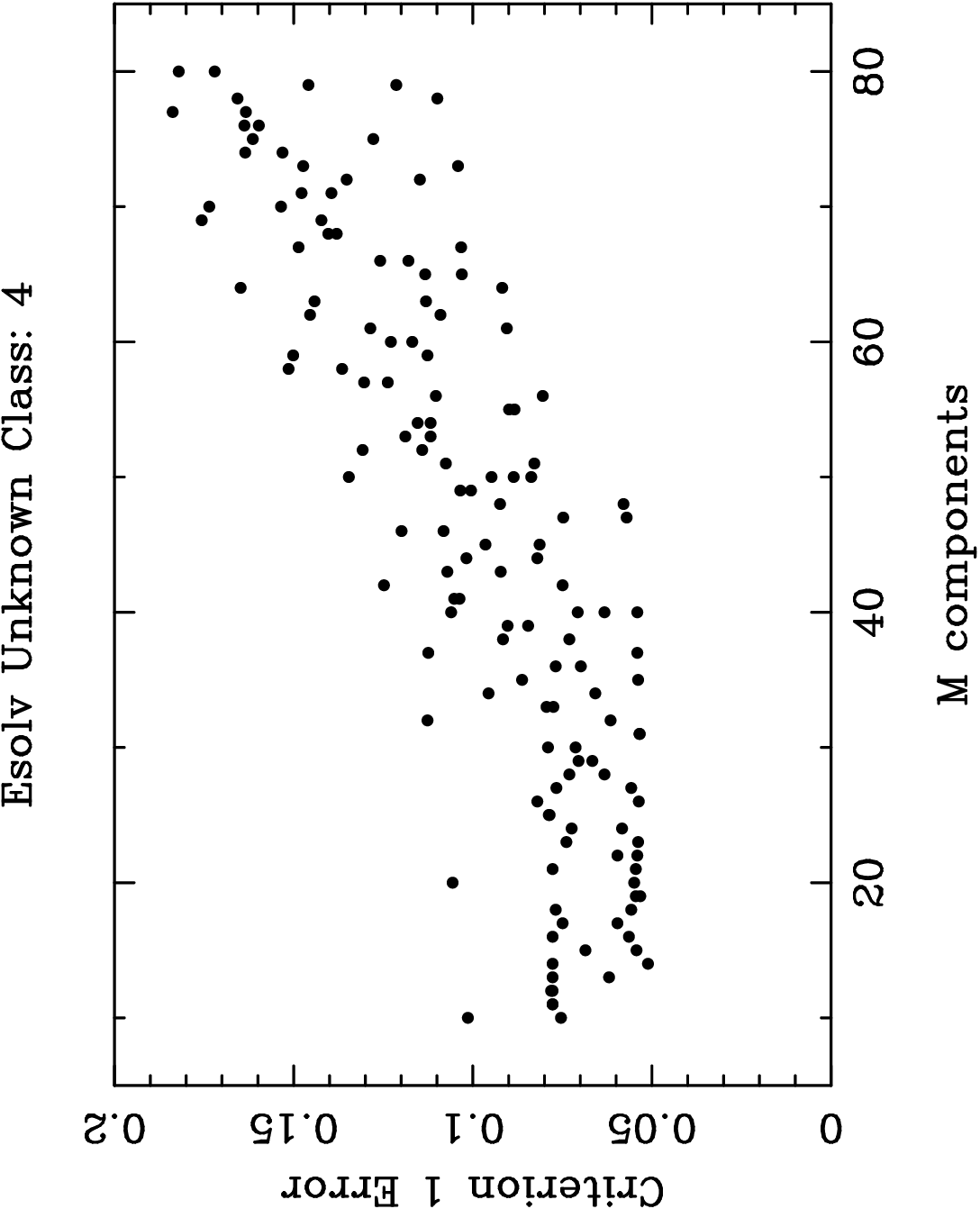}
\label{fig:fig3}
\caption{\small Criterion 1 error as a function of the number of
  components in the mixture model using ESOLV data with unknown class
  4.}
\end{figure}

\begin{figure}
\epsscale{0.8}
\plotone{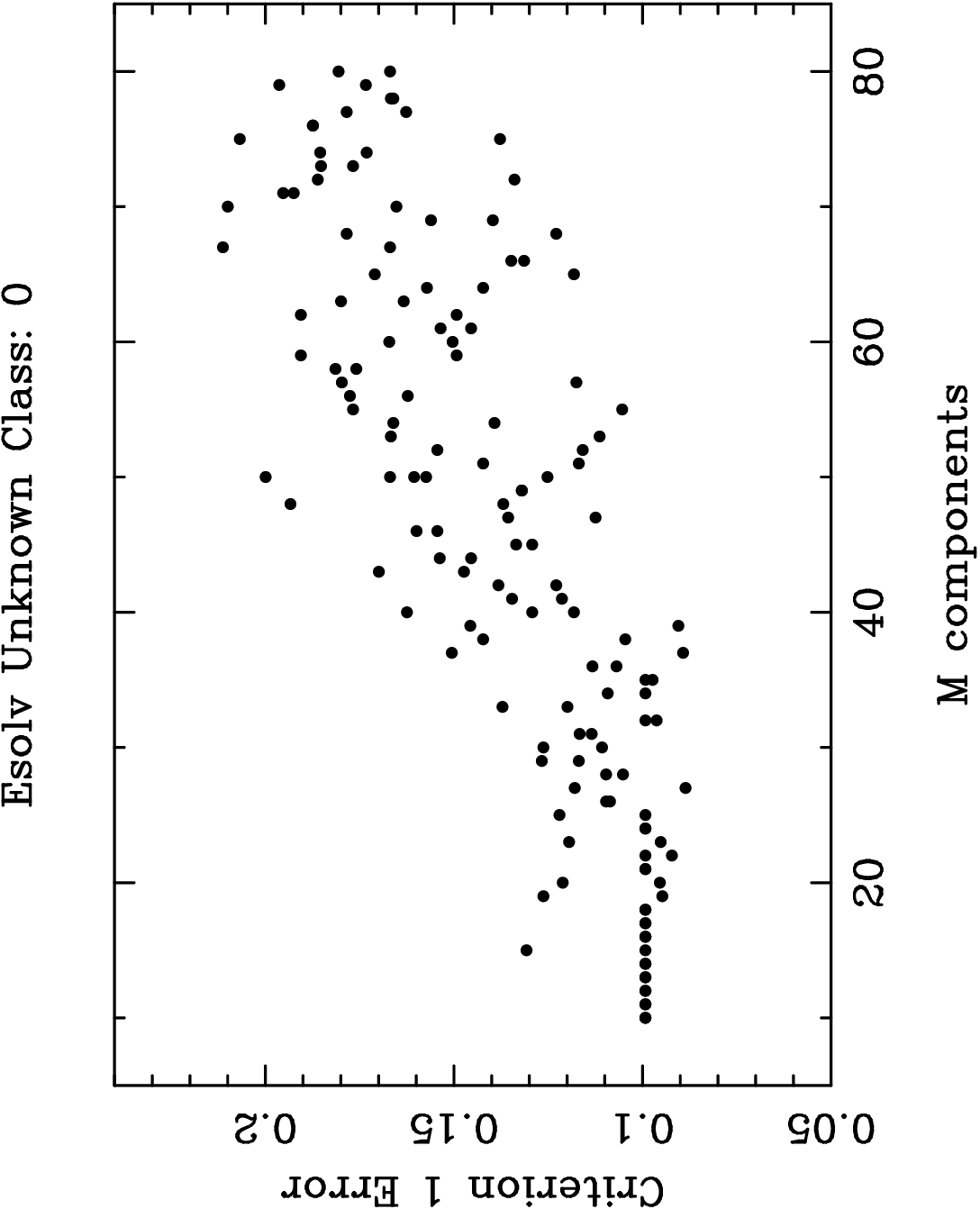}
\label{fig:fig4}
\caption{\small Criterion 1 error as a function of the number of
  components in the mixture model using ESOLV data with unknown class
  0.}
\end{figure}

Figures 5 and 6 show the Criterion 1 error as a function of the number
of mixture model components for the SDSS data.  Again we see the
Criterion 1 error remaining pretty much static until the model order
$M$ reaches 25 to 35 components.  At this point nonpredefined
components are added to the model allowing further decrease in the
Criterion 1 error.

\begin{figure}
\epsscale{0.8}
\plotone{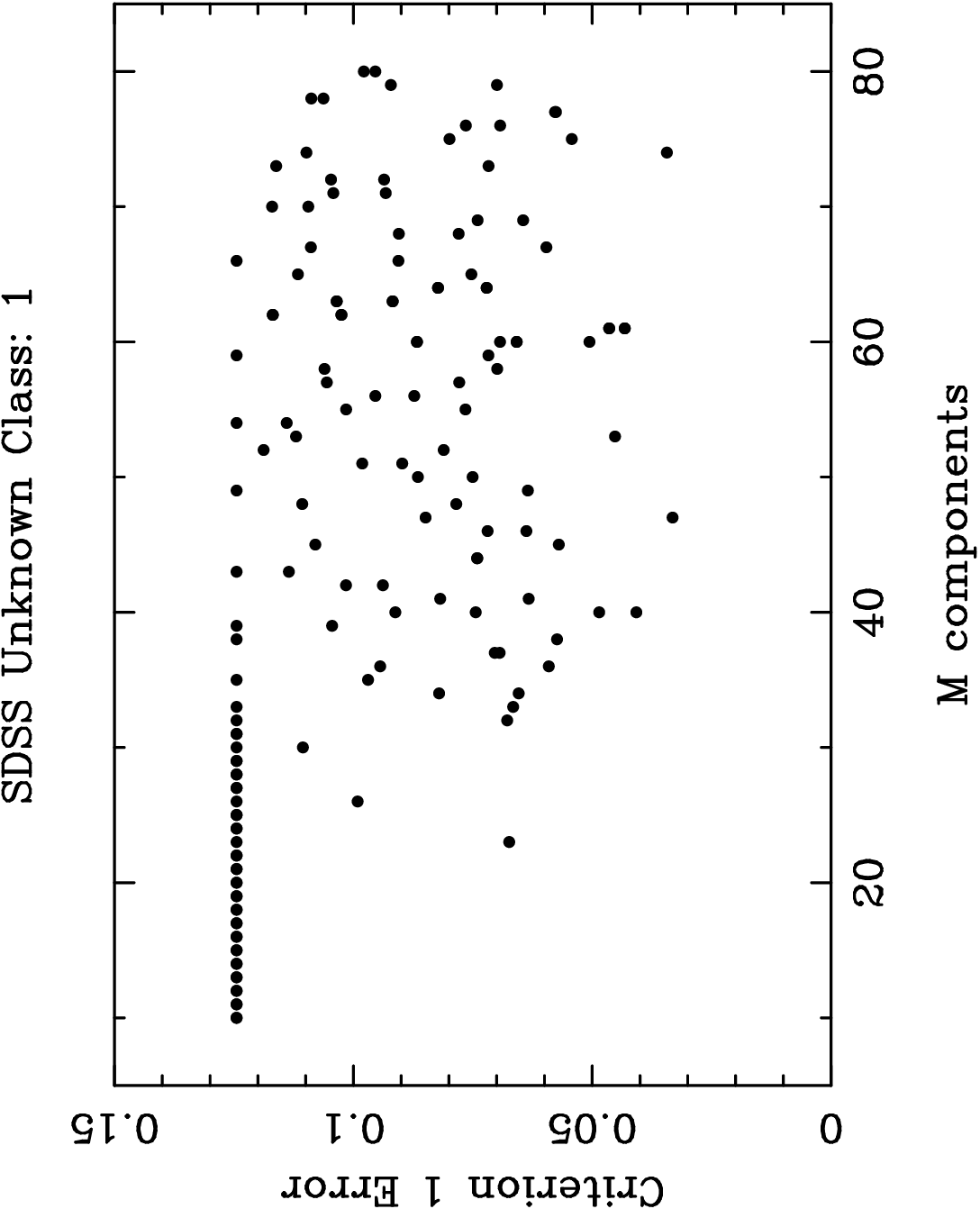}
\label{fig:fig5}
\caption{\small Criterion 1 error as a function of the number of
  components in the mixture model using SDSS data with unknown class
  1.}
\end{figure}

\begin{figure}
\epsscale{0.8}
\plotone{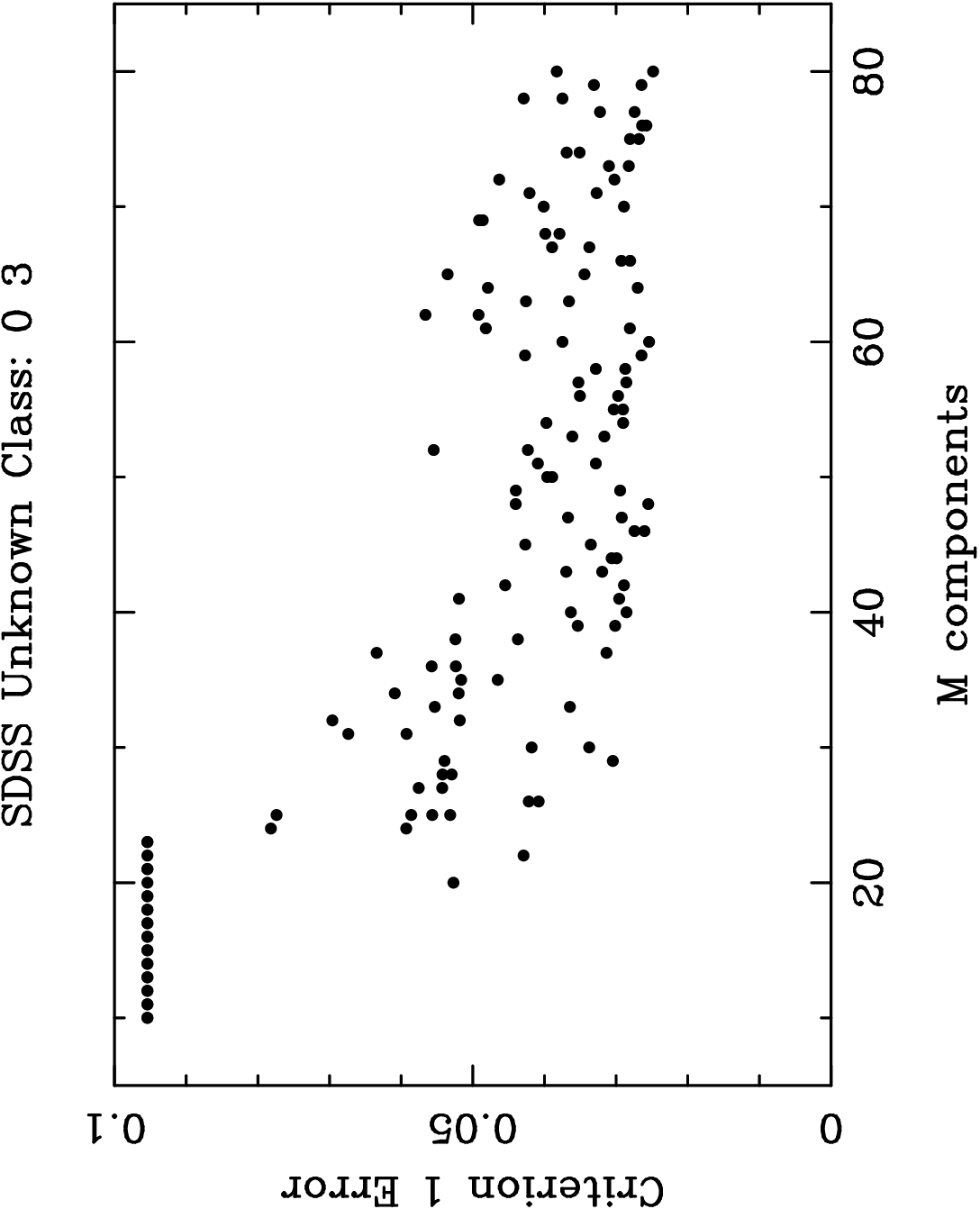}
\label{fig:fig6}
\caption{\small Criterion 1 error as a function of the number of
  components in the mixture model using SDSS data with unknown classes
  0 and 3.}
\end{figure}

The SDSS data contains approximately ten times the number of data
points as the ESOLV data.  The mixture model approach generally
requires a larger number of components to describe the SDSS data set.
On average 34 components were required to describe the ESOLV data
using the best model by Criterion 1, while 61 components were needed
to describe the SDSS data.  Using BIC to determine the best model
required on average 61 components for the ESOLV data and 71 for the
SDSS data.

\begin{deluxetable}{ccclllllr}
\tabletypesize{\scriptsize}
\tablecaption{Mixture model and neural network results using Criterion
  1 error model selection for ESOLV data.\label{tab:etf}}
\tablewidth{0pt}
\tablehead{
\colhead{} & \multicolumn{5}{c}{Mixture Model} & \colhead{} &
\colhead{Neural Network} & \colhead{} \\
\cline{2-6} \cline{8-8}\\
\colhead{Unknown Class} & \colhead{ncomp} & \colhead{nonpre} &
\colhead{Criterion 1} & \colhead{Criterion 2} &
\colhead{Criterion 3} & \colhead{} &
\colhead{Criterion 1} & \colhead{$\Delta \%$}}
\startdata
0 & 27 & 1   & 0.08859 & 0.7275 & 0.3742 && 0.09923 & -10.7\\
1 & 42 & 1   & 0.1717  & 0.9036 & 0.3220 && 0.1810  & -5.1\\
2 & 70 & 16  & 0.3797  & 0.5510 & 0.3083 && 0.4883  & -22.2\\
3 & 48 & 5   & 0.2183  & 0.7482 & 0.3249 && 0.2408  & -9.3\\
4 & 14 & 1   & 0.05110 & 0.5178 & 0.4221 && 0.07773 & -34.3\\
0 1 & 16 & 4 & 0.1399  & 0.5588 & 0.3276 && 0.2805  & -50.3\\
0 2 & 13 & 3 & 0.4553  & 0.6466 & 0.2140 && 0.6109  & -25.5\\
0 3 & 39 & 4 & 0.3179  & 0.8479 & 0.2524 && 0.3403  & -6.6\\
0 4 & 34 & 2 & 0.1463  & 0.7702 & 0.3573 && 0.1770  & -17.3\\
\cline{2-9}
    &$<33.7>$&$<4.1>$&$<0.2188>$&$<0.6978>$&$<0.3230>$&&$<0.2773>$&$<-20.2>$\\
\enddata
\end{deluxetable}

\begin{deluxetable}{ccclllllr}
\tabletypesize{\scriptsize}
\tablecaption{Mixture model and neural network results using
  Criterion 1 error model selection for SDSS data. \label{tab:stf}}
\tablewidth{0pt}
\tablehead{
\colhead{} & \multicolumn{5}{c}{Mixture Model} & \colhead{} &
 \colhead{Neural Network} & \colhead{} \\
\cline{2-6} \cline{8-8}\\
\colhead{Unknown Class} & \colhead{ncomp} & \colhead{nonpre} &
\colhead{Criterion 1} & \colhead{Criterion 2} &
\colhead{Criterion 3} & \colhead{} &
\colhead{Criterion 1} & \colhead{$\Delta \%$}}
\startdata
0 & 49 & 0 & 0.004711 & NA    & 0.06145 && 0.004711 & 0\\
1 & 47 & 5 & 0.03319  & 0.2091   & 0.04596 && 0.1244   & -73.3\\
2 & 28 & 20& 0.03377  & 0.006797 & 0.1117  && 0.1344   & -74.9\\
3 & 69 & 4 & 0.02181  & 0.1703   & 0.007064&& 0.09071  & -76.0\\
4 & 47 & 0 & 0.004876 & NA    & 0.06421 && 0.004876 & 0\\
5 & 68 & 0 & 0.002675 & NA    & 0.06290 && 0.002675 & 0\\
6 & 74 & 1 & 0.01187  & 0.3382   & 0.08020 && 0.02105  & -43.6\\
0 1 & 79 & 5 & 0.04074  & 0.2434 & 0.03501 && 0.1292   & -68.5\\
0 2 & 33 & 20 & 0.03393 & 0.01153 & 0.09026 && 0.8673  & -96.1\\
0 3 & 80 & 5 & 0.02483  & 0.2059  & 0.06232 && 0.09542 & -73.9\\
0 4 & 66 & 1 & 0.009176 & NA   & 0.07416 && 0.009587& -4.3\\
0 5 & 77 & 1 & 0.005719 & 0.6518  & 0.09580 && 0.007386& -22.6\\
0 6 & 76 & 1 & 0.01559  & 0.4553  & 0.06439 && 0.02576 & -39.5\\
\cline{2-9}
    &$<61>$&$<4.8>$&$<0.01868>$&$<0.2547>$&$<0.06580>$&&$<0.1167>$&$<-57.3>$\\
\enddata
\end{deluxetable}

\begin{deluxetable}{ccclllllr}
\tabletypesize{\scriptsize}
\tablecaption{Mixture model selected by BIC and neural network results
  on the ESOLV data. \label{tab:emdl}}
\tablewidth{0pt}
\tablehead{
\colhead{} & \multicolumn{5}{c}{Mixture Model} & \colhead{} &
 \colhead{Neural Network} & \colhead{} \\
\cline{2-6} \cline{8-8}\\
\colhead{Unknown Class} & \colhead{ncomp} & \colhead{nonpre} &
\colhead{Criterion 1} & \colhead{Criterion 2} &
\colhead{Criterion 3} & \colhead{} & 
\colhead{Criterion 1} & \colhead{$\Delta \%$}}
\startdata
0 & 60 & 7   & 0.1530 & 0.9206 & 0.3991 && 0.09923 &  54.2\\
1 & 53 & 10  & 0.2619 & 0.8672 & 0.3345 && 0.1810  &  44.7\\
2 & 63 & 11  & 0.4327 & 0.7428 & 0.3149 && 0.4883  & -11.4\\
3 & 54 & 6   & 0.2564 & 0.8896 & 0.3067 && 0.2408  &   6.5\\
4 & 67 & 7   & 0.1033 & 0.6137 & 0.4373 && 0.07773 &  32.9\\
0 1 & 69 & 13& 0.1902 & 0.6203 & 0.3184 && 0.2805  & -32.2\\
0 2 & 57 & 8 & 0.5253 & 0.7985 & 0.2091 && 0.6109  & -14.0\\
0 3 & 62 & 11& 0.3394 & 0.8542 & 0.2369 && 0.3403  &  -0.3\\
0 4 & 62 & 7 & 0.2004 & 0.7726 & 0.3684 && 0.1770  &  13.2\\
\cline{2-9}
    &$<60.8>$&$<8.9>$&$<0.2736>$&$<0.7866>$&$<0.3250>$&&$<0.2773>$&$<10.4>$\\
\enddata
\end{deluxetable}

\begin{deluxetable}{ccclllllr}
\tabletypesize{\scriptsize}
\tablecaption{Mixture model selected by BIC and neural network results
  on the SDSS data.\label{tab:smdl}}
\tablewidth{0pt}
\tablehead{
\colhead{} & \multicolumn{5}{c}{Mixture Model} & \colhead{} &
\colhead{Neural Network} & \colhead{}\\
\cline{2-6} \cline{8-8}\\
\colhead{Unknown Class} & \colhead{ncomp} & \colhead{nonpre} &
\colhead{Criterion 1} & \colhead{Criterion 2} &
\colhead{Criterion 3} & \colhead{} &
\colhead{Criterion 1} & \colhead{$\Delta \%$}}
\startdata
0 & 70 & 0  & 0.004711  & NA  & 0.1100  && 0.004711 &   0\\
1 & 63 & 5  & 0.1044   & 0.5353  & 0.03800 && 0.1244   & -16.1\\
2 & 67 & 24 & 0.09999  & 0.03841 & 0.09737 && 0.1344   & -25.7\\
3 & 74 & 5  & 0.02273  & 0.06070  & 0.1053  && 0.09071 & -74.9\\
4 & 76 & 0  & 0.004876 & NA   & 0.1060  && 0.004876 &   0\\
5 & 80 & 1  & 0.002921 & 0.5389  & 0.1178  && 0.002675 &   9.2\\
6 & 71 & 3  & 0.02732  & 0.6661  & 0.1066  && 0.02105  &  29.8\\
0 1 & 66 & 5 & 0.09434  & 0.4886 & 0.03388 && 0.1292   & -27.0\\
0 2 & 79 & 30 & 0.1128 & 0.02745 & 0.06373  && 0.8673  & -87.0\\
0 3 & 74 & 5 & 0.03306  & 0.1450 & 0.10390 && 0.09542  & -65.4\\
0 4 & 61 & 1 & 0.009505 & 0.5297  & 0.1121  && 0.009587&  -0.9\\
0 5 & 71 & 1 & 0.006275 & 0.6044  & 0.1116  && 0.007386& -15.0\\
0 6 & 71 & 3 & 0.02732  & 0.6661  & 0.1066 && 0.02576  &   6.1\\
\cline{2-9}
    &$<71>$&$<6.4>$&$<0.04233>$&$<0.3910>$&$<0.09330>$&&$<0.1167>$&$<-20.5>$\\
\enddata
\end{deluxetable}

\section{Discussion}
As can be seen from the results presented above we are, in general,
able to achieve a significantly lower Criterion 1 error value when
using unlabeled data to augment the labeled data.  Overall, we
obtained a $22\%$ (listed in table \ref{tab:etf} as the fraction
$0.2188$ not as the percent) Criterion 1 error for ESOLV data and a
$2\%$ error for SDSS data when using Criterion 1 as the model
selection method.  When using BIC for model selection we obtained
$27\%$ error for ESOLV data and $4\%$ error for SDSS data.  The
percentage change column of Table \ref{tab:etf} shows that on average
the mixture models reduced the Criterion 1 error for ESOLV data by
$20\%$, but this reduction was as high as $50\%$ when classes 0 and 1
were unknown.  Table \ref{tab:stf} similarly shows an average $57\%$
reduction in Criterion 1 error for SDSS data with a $96\%$ reduction
when classes 0 and 2 were unknown.  A $20\%$ reduction in error for
the ESOLV data implies an extra 1043 out of 5217 objects were
correctly classified by the mixture model compared to the neural
network.  For the SDSS data a $57\%$ error reduction implies an
additional 30784 out of 54007 objects were correctly classified by the
mixture model.

Our study is the first to apply semisupervised learning to
astronomical data, and the first, to our knowledge, to use a data set
as large as 50000 points.  This is an important test of the
methodology because of the vast amount of astronomical data freely
available today, most of which is unlabeled.  Demonstrating that our
methods work with large astronomical data sets was a primary goal of
this work.

Nevertheless, there are a number of factors that influence the
reliability of the proposed method and what level of error can be
achieved.  The results in Tables 3-6 demonstrate the importance of the
model order selection technique.  BIC-based selection fares well on
the SDSS data, achieving substantially better average Criterion 1
results than the neural network optimized for Criterion 1 and only
modestly worse results than the mixture model optimized for Criterion
1 (0.02 vs. 0.04 average error rates).  However, there is a
significant average performance gap between the two mixture approaches
on the ESOLV data (0.22 vs. 0.27) and the BIC-selected mixture is only
comparable to the neural network on ESOLV (0.273 vs. 0.277 average
error rates).  It is possible that a better model order selection
technique could improve the mixture results on ESOLV.

One artifact of optimizing the mixture for Criterion 1 is that, on
average, smaller models are selected, compared with the mixtures
selected by BIC.  For ESOLV on average 34 components were selected by
the former approach, while on average 61 components were selected by
the latter.  Furthermore, an average of 4.1 nonpredefined components
were used with Criterion 1 model selection while an average of 8.9
were used with BIC.  For the SDSS data an average of 61 components,
including 4.8 nonpredefined components, were selected using Criterion
1.  Finding the best SDSS model by BIC we needed on average 71
components, including 6.4 nonpredefined components.  While we learned
models with up to 80 components, in some cases for the SDSS data the
best models were using close to 80 components.  This suggests it may
be reasonable to evaluate solutions with even more components.  While
the mean number of components selected by BIC was greater than that
selected according to Criterion 1, the variance in the number of
selected components is much greater for selection according to
Criterion 1.  This is consistent with the results in Figure 2, which
indicate that, for best Criterion 1 performance, the number of
components is significantly correlated with the mass of the unknown
classes (which varies greatly since the classes are far from equally
likely).  This further means that, in some cases, when the mass of the
unknown classes is large, Criterion 1 selects more components than
BIC.  For example, for the ESOLV data, class 2 occurs $46\%$ of the
time.  When this class is taken as unknown, BIC selected 63
components, while Criterion 1 selected 70.

Another factor that influences model accuracy is the fact that the
learning objective function ${\cal L}$ is multimodal, with significant
potential for finding suboptimal local maxima, rather than the global
maximum.  At each model order, we generated several solutions based on
different initializations and picked the one with greatest
log-likelihood.  However, there is anecdotal evidence in our results
that we may only be finding locally optimal solutions at each model
order.  Referring back to Figure 2 we see that the best
model for unknown classes 0 2 contains only 13 mixture components.
This clearly is not in keeping with the trend that more mixture
components are needed to explain the data with larger mass fraction of
unknown classes.  It appears in this case that a particularly good
solution was found.  This likewise suggests that, at other orders,
suboptimal, local maximum solutions were found.

The Criterion 2 error is a measure of how well the algorithm can
classify objects within the newly found classes.  This is a very hard
problem because we are asking the algorithm to do two things.  First,
determine if some mixture components (the nonpredefined components)
are needed to describe objects that do not fit into the existing known
class structure.  Second, partition these objects correctly between
the unknown class components.  This second step is effectively looking
for substructure in the newly discovered classes.

For the ESOLV data we found on average about a $70\%$ Criterion 2
error compared with about a $25\%$ error for the SDSS data when using
Criterion 1 model selection.  Similarly, we find about $79\%$ error
for the ESOLV data and $39\%$ error for SDSS data when using BIC model
selection.  The values of NA for Criterion 2 error for some of the
SDSS experiments reflect models where there were no nonpredefined
components; in this case the error measure is undefined.  Note that,
in all these cases, the unknown classes had very small mass, which
explains why no nonpredefined components were found.  In particular,
class 0 and class 4 collectively comprise less than $1\%$ of the SDSS
data.  Thus, when these classes are missing, we would not expect to
find nonpredefined components in the solution unless both 1) there are
more than 100 components in the model and 2) the model criterion
selects a solution of this size.

The Criterion 3 error measures how well the model can assign objects
to known classes.  For this we obtained about a $32\%$ error for ESOLV
data and $7\%$ error for SDSS data using Criterion 1 model selection.
When using BIC model selection we obtained $33\%$ error for ESOLV data
and $9\%$ error for SDSS data.

We find the overall results presented here very promising.  The tests
done here have demonstrated the efficacy of the class discovery
problem and approaches.  However, more work will be required to
develop a mature technology for highly reliable new class discovery.

\section{Acknowledgments}
We would like to thank the NASA Applied Information Systems Research
Program for supporting us in this effort under contract NAS5-02098.
One of the authors (DB) would like to thank Ofer Lahav for supplying
the ESO-LV data.

\end{document}